# Efficient surface second-harmonic generation in slot micro/nano-fibers

Wei Luo, Fei Xu* and Yan-qing Lu*

*National Laboratory of Solid State Microstructures and College of Engineering and Applied Sciences, Nanjing University, Nanjing 210093, P. R. China*

[*]*Corresponding author: feixu@nju.edu.cn and yqlu@nju.edu.cn*

We propose to use slot micro/nano-fiber (SMNF) to enhance the second-harmonic generation based on surface dipole nonlinearity. The slot structure is simple and promising to manufacture with high accuracy and reliability by mature micromachining techniques. Light field can be enhanced and confined, and the surface area can be increased in the sub-wavelength low-refractive-index air slot. The maximum conversion efficiency of the SMNFs in our calculations is about 24 times higher than that of circular micro/nano-fibers. It is promising to provide a competing platform for a new class of fiber-based ultra-tiny light sources spanning the UV- to the mid-infrared spectrum.

PACS numbers: 42.65.Ky, 73.20.Mf, 78.20.Bh

# 1. Introduction

Nonlinear interactions in optical fibers have been extensively studied since the early 1970s. Because of the central symmetry that a doped-silica fiber is supposed to have, all the second-order dipole nonlinear coefficients should be zero, and second-order nonlinearities should not appear in silica optical fibers. Nevertheless, second-harmonic generation (SHG) with peak power-conversion efficiency as high as ~3% has been reported to occur in optical fibers in 1986 [1]. Initially, this phenomenon could not be reasonably explained by core-cladding interface or bulk multipole moment contributions to the second-order nonlinearity [2]. A subsequent study showed that the nonlinearity caused by the formation of a second-order susceptibility ($\chi^{(2)}$) grating through multiphoton processes involving both pump and SHG light agreed with the experimental results [3]. The $\chi^{(2)}$ grating introduces the second-order polarization and compensates the phase mismatch arising from waveguide and material dispersion in the fiber. However, it has proved impossible to improve the SHG conversion efficiency beyond the level of a few percent because of a self-saturation effect by the interference of the SHG light with the $\chi^{(2)}$ grating [3].

A recent experiment found phase-matched SHG at 532nm in a low-order mode of a sub-micron diameter glass fiber [4]. The multiphoton processes leading to the $\chi^{(2)}$ grating are limited in silica microfibers for the weak photosensitivity. In fact, a submicrometric diameter of microfibers calls for reexamination of interface and bulk multipole moment contributions to $\chi^{(2)}$. Sub-micron diameter silica fiber can provide a higher power density, so that surface nonlinearity of core-cladding interface and nonlinearity of bulk multipole become the major mechanism for SHG. Furthermore, large core-cladding index contrast makes it possible to achieve SHG phase matching in a low-order mode with a sufficient intensity at the surface.

The microstructured optical fibers developed during the last decade offer a lot of new features. With micromachining technologies such as focused ion beam (FIB) milling, different geometry can be obtained in optic fibers, for example, Fabry-Perot cavity with an open notch in a circular microfiber[5, 6], ultra-short Bragg grating with deep grooves [7, 8], fiber-top cantilever [9], and sub-wavelength light confinement tip [10]. Combining fiber-drawing and micromachining technologies, a slot structure is possible to realize in the waist of the circular micro/nano-fiber (CMNF), creating the so-called slot micro/nano-fiber (SMNF), which introduces a high birefringence and a great power density around the slot [11]. The high intensity at the slot surface helps to enhance the surface nonlinearity.

In this letter, we investigate the surface SHG by phase matching between fundamental and low-order mode in SMNFs, and compare the results with that of CMNFs. CMNFs are studied by analytical methods, and SMNFs will be studied numerically using the finite-element method. It can be seen that higher SHG conversion will theoretically be achieved in SMNFs. The maximum conversion efficiency in our calculations is about 24 times larger than that of CMNFs. This kind of nano-scale geometry should open up new possibilities in fiber functionality including fiber-based optical nanosource, as well as nonlinear signal proceeding.

## 2. Theory analysis and numerical model

In the small-signal limit (pump depletion is negligible), SHG process can be described by the equation of amplitudes coupling [12]:

$$\frac{dA_2}{dz} - i\rho_2 A_1^2 \exp(i\Delta\beta z) = 0 \quad (1)$$

where $A_1$ and $A_2$ are the field amplitudes of the pump and SHG signals, respectively; $\Delta\beta=2\beta_1-\beta_2$ is the phase mismatch between the fundamental and second-harmonic waves; $\rho_2$ is the overlap integral [2]:

$$\rho_2 = \frac{\omega_2}{4N_1\sqrt{N_2}} \int \mathbf{e}_2^* \cdot \mathbf{P}^{(2)} dS \quad (2)$$

where $\omega_2$ is the second-harmonic angular frequency. The function will be integrated on the cross-sectional region of the fiber, and $dS$ is the area element. All the field components are normalized by the normalizing factors:

$$N_j = \frac{1}{2}\int |(\mathbf{e}_j^* \times \mathbf{h}_j)\cdot \hat{\mathbf{z}}| dS \quad (j=1,2) \quad (3)$$

The fields of the guided modes can be written as:

$$\mathbf{E}(\mathbf{r},\omega_j) = A_j(\omega_j)\mathbf{e}_j(\mathbf{r},\omega_j)\exp(i(\beta_j z - \omega_j t)) \quad (4)$$

$$\mathbf{H}(\mathbf{r},\omega_j) = A_j(\omega_j)\mathbf{h}_j(\mathbf{r},\omega_j)\exp(i(\beta_j z - \omega_j t)) \quad (5)$$

$\mathbf{P}^{(2)}$ is the second-order nonlinear polarization. For pure-silica microfibers in air cladding, it originates from the contributions of silica-air interface and bulk multipole moments. The bulk contributions are expressed as [12, 13]:

$$\mathbf{P}_b^{(2)}(\mathbf{r}) = \varepsilon_0 \gamma \nabla(\mathbf{E}_1 \cdot \mathbf{E}_1) + \varepsilon_0 \delta(\mathbf{E}_1 \cdot \nabla)\mathbf{E}_1 \quad (6)$$

A third term proportional to $\mathbf{E}_1(\nabla \cdot \mathbf{E}_1)$ can be included in the surface term [2]. The bulk contributions shown in eq. (6) can be ignored, since the results for the silica microfibers indicated them to be of minor importance [12]. So we just take surface contributions into account, and $\mathbf{P}^{(2)}$ can be written as:

$$\mathbf{P}^{(2)} \approx \mathbf{P}_s^{(2)}(\mathbf{r}) = \delta(\mathbf{r}-\mathbf{S})[\mathbf{P}_\perp^{(2s)} + \mathbf{P}_{\perp\parallel}^{(2s)} + \mathbf{P}_\parallel^{(2s)}] \quad (7)$$

where **S** stands for the vectors of the silica-air interface. The surface contributions can be divided into three distinct terms [12]:

$$\mathbf{P}_\perp^{(2s)} = \varepsilon_0 \chi_\perp^{(2s)} \mathbf{e}_{1\perp}^2 \hat{\mathbf{r}}_\perp \quad (8)$$

$$\mathbf{P}_{\perp\parallel}^{(2s)} = \varepsilon_0 \chi_{\perp\parallel}^{(2s)} \mathbf{e}_{1\parallel}^2 \hat{\mathbf{r}}_\perp \quad (9)$$

$$\mathbf{P}_\parallel^{(2s)} = 2\varepsilon_0 \chi_\parallel^{(2s)} e_{1\perp} \mathbf{e}_{1\parallel} \quad (10)$$

where $\hat{\mathbf{r}}_\perp$ is the unit vector normal to the interface. The three terms of surface second-order susceptibility can be measured by experiments: $\chi_\perp^{(2s)} = 6.3 \times 10^3 \, \text{pm}^2/\text{V}$, $\chi_{\perp\parallel}^{(2s)} = 7.7 \times 10^2 \, \text{pm}^2/\text{V}$, $\chi_\parallel^{(2s)} = 7.9 \times 10^2 \, \text{pm}^2/\text{V}$ [12, 14, 15].

For CMNF, the fields of the guided modes are available by solving the Maxwell equations [16]. In our calculation, the pump signal is assumed to propagate in the $HE_{11}$ mode, while the SHG signal is assumed to generate in the $HE_{21}$ mode. The phase matching is achieved by material dispersion and multimode dispersion of the fiber. The material dispersion of silica glass can be described by the Sellmeier polynomial [16]. For $HE_{n1}$ mode, the propagation constant $\beta_n$ is determined from the equation:

$$\left[\frac{J_n'(u_n)}{u_n J_n(u_n)} + \frac{K_n'(w_n)}{w_n K_n(w_n)}\right]\left[\frac{J_n'(u_n)}{u_n J_n(u_n)} + \frac{1}{n_s^2}\frac{K_n'(w_n)}{w_n K_n(w_n)}\right] = n^2\left(\frac{1}{u_n^2} + \frac{1}{w_n^2}\right)\left[\frac{1}{u_n^2} + \frac{1}{(n_s w_n)^2}\right] \quad (11)$$

$$u_n = ak_n\sqrt{n_s^2 - n_n^2}, \quad w_n = ak_n\sqrt{n_n^2 - 1} \quad (12)$$

where $J_n$ is Bessel function of the first kind, $K_n$ is modified Bessel function of the second kind, subscript n is the mode order, $k_n$ is the vacuum wave vector of the guided light, $n_s$ is the silica refractive index calculated from the Sellmeier polynomial, $n_n = \beta_n/k_n$ is the modal effective index, and a is the fiber radius. SHG signal will only efficiently generate in the phase-matched mode with $\Delta\beta=2\beta_1-\beta_2=0$. In order to realize phase matching in different SHG frequencies, we modify the multimode dispersion by changing the fiber diameter.

In the small-signal limit with perfect phase matching, the power-conversion efficiency is given as [2]:

$$\frac{P_2}{P_1}=(\rho_2 z)^2 P_1 \qquad (13)$$

where $P_1$ is the pump power and z is the interaction length (generally the waist length of microfiber or the slot length). This is not a complete description of SHG dynamics, but it is sufficient to estimate the SHG conversion efficiencies for comparison between CMNFs and SMNFs.

According to eq. (13), SHG conversion efficiency is proportional to the square of the overlap integral $\rho_2$ with all other conditions being equal. Thus the absolute value of $\rho_2$ determines the SHG conversion capability. From eq. (2) – (12) and analytic solutions of mode fields, we can calculate the $|\rho_2|$ of CMNFs.

For SMNFs, numerical simulations will be adopted to obtain $|\rho_2|$. In our work, mode fields and propagation constants are determined using the finite-element method. The SMNF structures are depicted in figure 1. There can be one or more slots in a microfiber. In the calculations, we just consider single-slot micro/nano-fibers (SSMNFs) and double-slot micro/nano-fibers (DSMNFs), and assume $n_{air}=1$. The parameters characterizing the slot structure are the diameter of the micro/nano-fiber d, the slot width $w_s$ ($w_{s1}$ and $w_{s2}$ for DSMNFs), the slot height $h_s$ ($h_{s1}$ and $h_{s2}$ for DSMNFs), and the distance between slots $d_s$ for DSMNFs. For SMNFs, the pump signal is assumed to propagate in the $HE_{11}$-like fundamental mode, while the SHG signal is assumed to generate in the $HE_{21}$-like mode (the 5th-order mode). $|\rho_2|$ of SMNFs is determined by eq. (2) – (10) and numerical solutions of mode fields and propagation constants.

## 3. Simulation results and discussions

Figure 2 shows the calculated relation between the fiber diameter and the phase-matched SHG wavelength for CMNF, SSMNF ($h_s$=0.5d, $w_s$=0.05d) and DSMNF ($h_{s1}$=$h_{s2}$=0.5d, $w_{s1}$=$w_{s2}$=0.05d, $d_s$=0.075d). As the slot number increases, phase-matched $\lambda_{SHG}$ for different structures at the same diameter decreases. It results from the modulation of the waveguide dispersion by slot structure. The slot structure enlarges the waveguide dispersion by more evanescent field propagating outside the fiber, making $\beta_1$ of the pump wave in fundamental mode and $\beta_2$ of the second-harmonic wave in high-order mode matched at a shorter wavelength. Modulation can be enhanced by more slots in the fiber.

In figure 3, the absolute value of $\rho_2$ for CMNF, SSMNF1 ($h_s$=0.5d, $w_s$=0.05d), SSMNF2 ($h_s$=d, $w_s$=0.05d), DSMNF1 ($h_{s1}$=$h_{s2}$=0.5d, $w_{s1}$=$w_{s2}$=0.05d, $d_s$=0.075d), and DSMNF2 ($h_{s1}$=$h_{s2}$=d, $w_{s1}$=$w_{s2}$=0.05d, $d_s$=0.075d) is plotted versus the phase-matched $\lambda_{SHG}$. $|\rho_2|$ roughly scales with $(\lambda_{SHG})^{-3}$ in all the five structures. $|\rho_2|$ of SMNFs is significantly larger than that of CMNF, and SSMNF2 has the maximum $|\rho_2|$ (about 5 times of that in CMNF). This can be explained by the increasing of the surface area and the power density at the surface. More surface area and higher surface power density contribute to stronger surface nonlinearity. Figure 4 shows the power flow distribution of CMNF, SSMNF, and DSMNF in $HE_{21}$ or $HE_{21}$-like mode for the corresponding phase-matched $\lambda_{SHG}$. In the figure, we can see a fraction of light field is confined in the slot structure and there is more evanescent field outside the SMNFs. Thus it enlarges the surface power density. However, the double slots make the power flow distribution dispersed, which decreases the light intensity at the surface. But they also increase the surface area at the same time. The surface area contribution is greater than the surface power density dispersion contribution in DSMNF1 and DSMNF2, so that DSMNFs have larger $|\rho_2|$ than SSMNF1. Surface area scales up with the height of the slots. With concentrated power distribution and large surface area in the slot, SSMNF2 has the maximum $|\rho_2|$ of all the fibers in calculations. Impacts of $h_s$ on $|\rho_2|$ can been seen more clearly by modulating $h_s$ of the SSMNF with $w_s$=0.05d (shown in figure 5).

Figure 6 shows the $|\rho_2|$-$\lambda_{SHG}$ relation with modulation of $w_s$ in the SSMNF with $h_s$=d. $|\rho_2|$ decreases as $w_s$ increases. A wider slot dispersed the power flow distribution, which leads to a lower light intensity at the surface and then the reduction of $|\rho_2|$. Further modulation of $d_s$ of the DSMNF with $h_{s1}$=$h_{s2}$=d and $w_{s1}$=$w_{s2}$=0.05d shows that the distance between slots has little influences on the overlap integral.

According to eq. (13), absolute value of $\rho_2$ determines the SHG conversion efficiency. From the simulation results, it proves that the SHG conversion efficiency achieved in SMNFs is significantly higher than in CMNFs. For SSMNF2 of slot length z=1mm and pumping peak power $P_1$=1kW with a 1550nm femtosecond laser source, conversion efficiency is calculated to be ~0.027%, while a CMNF with the same parameters only has conversion efficiency ~0.0011%. Further improvement of the SHG efficiency can be realized by optimizing the structural parameters and utilizing other mechanism of second-order nonlinearity. For example, strain in the material breaks the symmetry of the structure, introducing a sizeable second-order nonlinearity into the waveguide, so that a stressing overlayer can be deposited on the fiber to enhance the nonlinearity [17].

## 4. Conclusions

In this work, the surface dipole contributions to the second-harmonic generation in slot microfibers have been studied numerically. According to our calculations, the introduction of the slot structure can significantly increase the surface second-order nonlinearity. Light field can be enhanced and confined, and the surface area can be increased in the sub-wavelength-wide low-refractive-index air slot. Two kinds of typical cases (SSMNF and DSMNF) are investigated and compared with CMNF. Surface area and surface power density are key factors to characterize the surface SHG conversion capability. The maximum $|\rho_2|$ in the calculations is about 5 times of that in CMNF, which equals to a SHG conversion efficiency about 24 times larger than that of CMNF. SMNFs can be fabricated by mature micromachining techniques such as FIB milling and femtosecond laser etching, and higher conversion efficiency is expected by the optimization of the SMNF parameters and other mechanism such as strain-induced second-order nonlinearity. The advantages of strong surface second-order nonlinearity, long interaction length and simple structure offer prospects for SMNFs in efficient SHG conversion applications. Its unique geometry can also provide a promising platform for ultra-small fiber laser in particular including ultra-violent and visible light.

**Acknowledgements**

This work is supported by National 973 program under contract No. 2012CB921803 and 2011CBA00205, National Science Fund for Distinguished Young Scholars under contract No. 61225026. The authors also acknowledge the support from PAPD and the Fundamental Research Funds for the Central Universities.

**Figure Captions:**

Fig. 1: Schematic of SMNF. Slot structure is located in the waist region, and slot number can be one or more. Inset, (a) Cross-sectional view of SSMNF in air cladding. $n_s$ and $n_{air}$ are the refractive index of silica and air, respectively. The waist diameter d, the slot width $w_s$ and the slot height $h_s$ characterize the structural features of the SSMNF. (b) Cross-sectional view of DSMNF in air cladding. Each slot has its own structural parameters ($w_{s1}$, $h_{s1}$ for left slot and $w_{s2}$, $h_{s2}$ for right slot). Parameter $d_s$ is the distance between the two slots.

Fig. 2: Phase-matched SHG wavelength $\lambda_{SHG}$ versus the fiber diameter d in CMNF (black line), SSMNF (red line) and DSMNF (blue line).

Fig. 3: Absolute value of overlap integral $|\rho_2|$ versus $\lambda_{SHG}$ for CMNF (orange line), SSMNF1 (green line), SSMNF2 (black line), DSMNF1 (blue line), and DSMNF2 (red line).

Fig. 4: Power flow distribution of CMNF (left), SSMNF (center), and DSMNF (right) in $HE_{21}$ or $HE_{21}$-like mode.

Fig. 5: The $|\rho_2|$-$\lambda_{SHG}$ relation for SSMNF with $h_s$=0 (blue line), 0.5d (green line), 0.75d (red line), and d (black line), respectively.

Fig. 6: Relation between $\lambda_{SHG}$ and $|\rho_2|$ for for SSMNF with $w_s$=0.1d (green line), 0.075d (red line), and 0.05d (black line), respectively.

Fig. 1

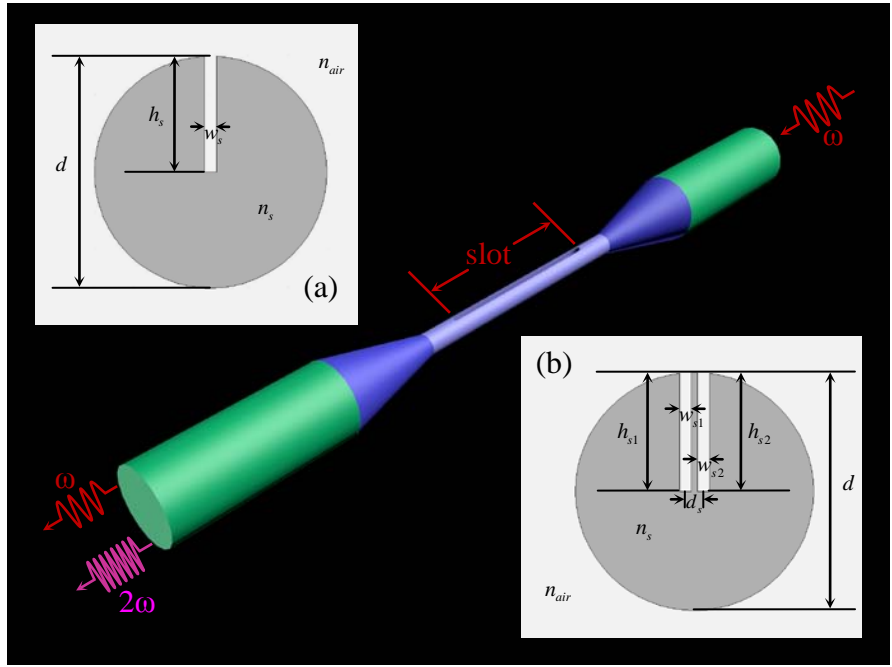

Fig. 2

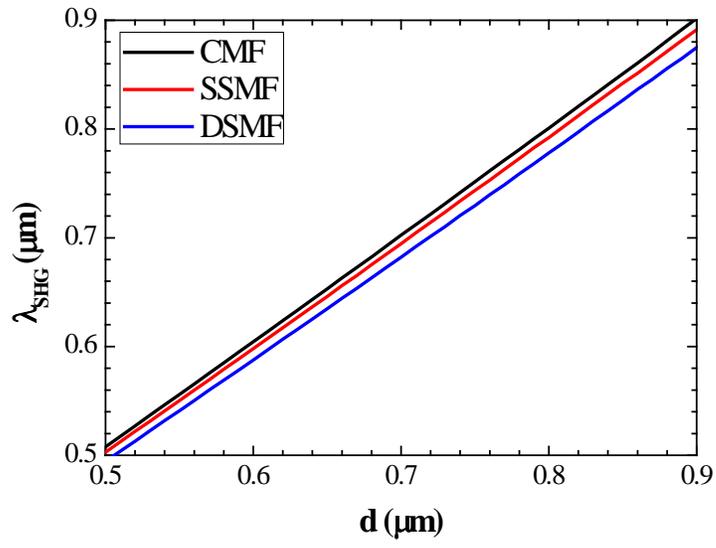

Fig. 3

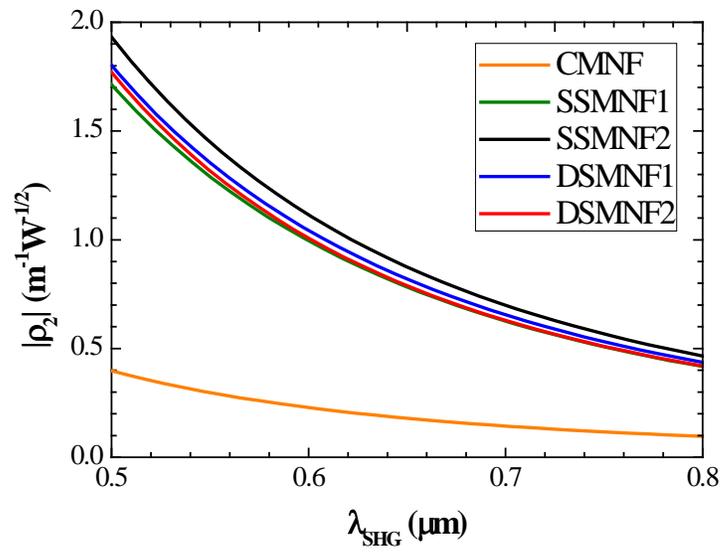

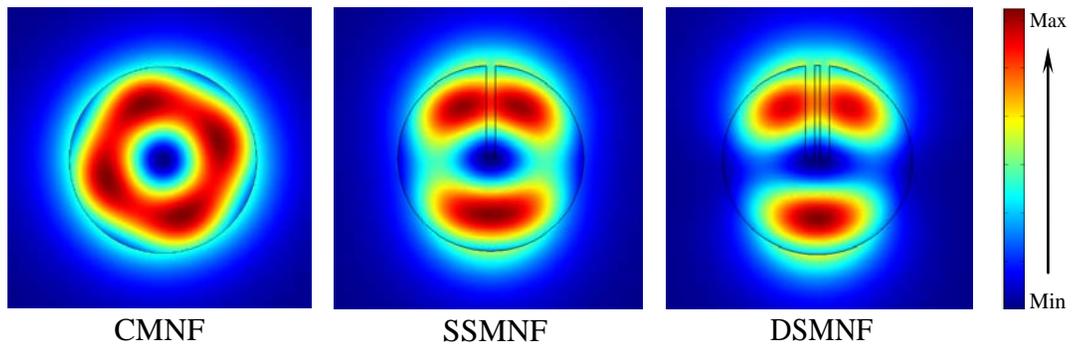

Fig. 5

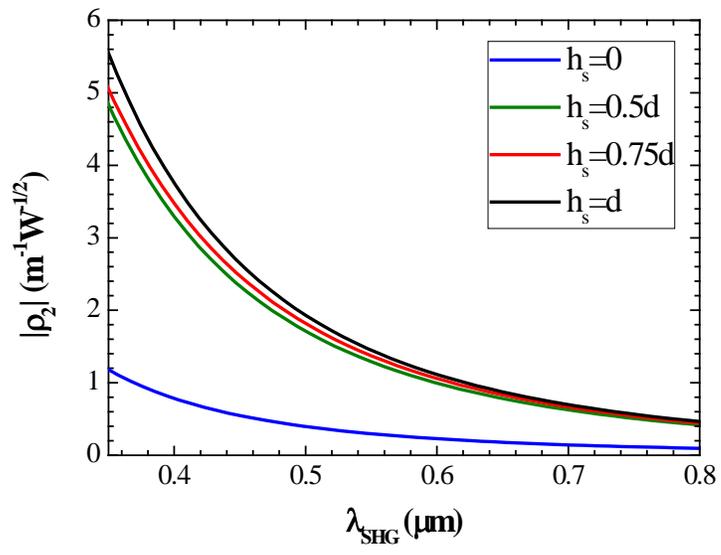

Fig. 6

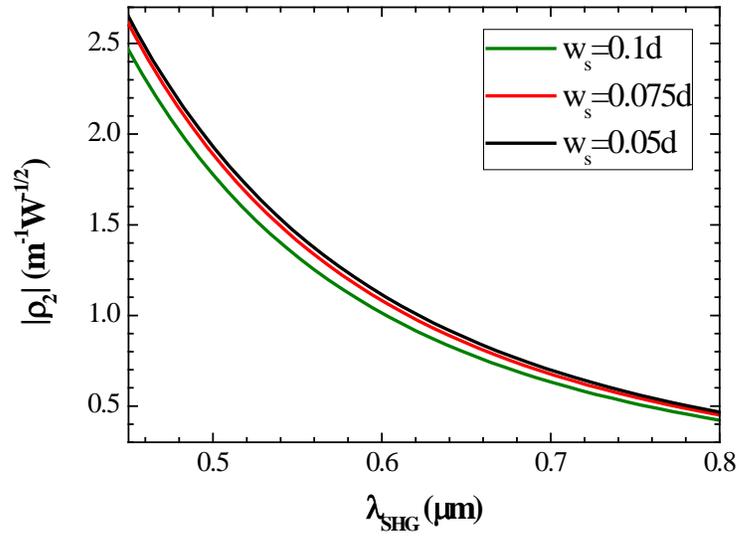